\documentclass{aa}  

\usepackage{graphicx}
\usepackage{txfonts}
\usepackage[colorlinks,citecolor=blue]{hyperref}
\usepackage{subfigure}
\usepackage{subcaption}

\begin{document} 

\title{3D particle-in-cell simulations of pulsar wind -- disk interaction: application to the transitional millisecond pulsar PSR J1023$+$0038}

\titlerunning{Pulsar wind -- disk interaction}

\author{Valentina Richard-Romei\inst{1}\and Benoît Cerutti\inst{1} \and Alessandro Papitto\inst{2} \and Riccardo La Placa\inst{2}}

\institute{Univ. Grenoble Alpes, CNRS, IPAG, 38000 Grenoble, France \and INAF, Osservatorio Astronomico di Roma \\
           \email{valentina.richard-romei@univ-grenoble-alpes.fr}\\
           \email{benoit.cerutti@univ-grenoble-alpes.fr}\\
           \email{alessandro.papitto@inaf.it}\\
           \email{riccardo.laplaca@inaf.it}
           }

\date{Received \today; accepted X}

 
\abstract
{Transitional millisecond pulsars constitute a peculiar subclass of neutron stars in which the pulsar alternates between accretion-powered and rotation-powered states, depending on the variations in the mass accretion flow coming from a low-mass companion star. A third intermediate state, referred to as ``sub-luminous disk state'', has been identified. During this state, observations indicate the presence of a disk surrounding the pulsar, and the system exhibits intriguing features, such as broad optical and X-ray pulsations characterized by a high luminosity. To date, no ab initio model of a pulsar wind interacting with an accretion disk has been developed to address these observables.}
{We aim to model the interaction of a pulsar wind with a surrounding disk in order to investigate the origin of the enhanced electromagnetic emission and to reproduce the observable properties of such systems, including their light curves and polarization signatures.}
{We perform three-dimensional particle-in-cell simulations of a pulsar magnetosphere surrounded by a perfectly conducting torus to model the interaction between the pulsar wind and the disk. We conduct a parametric study to assess the impact of the pulsar's magnetic obliquity.}
{We find that the presence of the disk induces a significant reconfiguration of the magnetosphere compared to the rotation-powered state, leading to enhanced plasma density at the inner disk boundary, increased magnetic field strength, and more efficient plasma isotropization and particle acceleration. As a result, the synchrotron radiation is substantially enhanced, and characterized by a strong continuous component and either one or two-peaked light curves, depending on the pulsar's magnetic obliquity. The polarization degree is reduced compared to isolated systems, and its energy dependence is explored. The rotation of the polarization angle can also be altered, depending on the observer's viewing angle.}
{The model successfully reproduces some of the main features of the optical and X-ray pulsed emission originating from PSR J1023$+$0038, thereby corroborating the scenario in which these pulsations originate from synchrotron radiation generated as the pulsar wind interacts with the inner edge of the disk.}


\maketitle

\section{Introduction}
Millisecond pulsars (MSPs), old neutron stars that display spin periods $< 30\,{\rm ms}$, are thought to be descendants of low-mass X-ray binaries (LMXBs), systems where matter from a low-mass donor star ($< 1 {\rm M}_\odot$) is accreted onto a neutron star, thereby transferring angular momentum and spinning up the pulsar to millisecond periods. Two main categories of millisecond pulsars can be distinguished: accreting millisecond pulsars (AMXPs; see \citealt{Campana_2018} for a review), a subclass of LMXBs, and rotation-powered millisecond pulsars, referred to as spider pulsars when the companion is a main-sequence star \citep{Roberts_2013}. However, since the early 2000s, a new class referred to as ``transitional millisecond pulsars'' (tMSPs) has been discovered, in which pulsars alternate between accretion-powered,  rotation-powered, and a third, intermediate state called ``sub-luminous disk state" (see \citealt{Campana_2018, Papitto_2022} for reviews). Three such sources have been identified so far \citep{Archibald_2009, Papitto_2013, Bassa_2014}. The sub-luminous disk state owes its name to its intermediate X-ray luminosity of $L_X\sim 10^{33}\text{--}10^{34}~\mathrm{erg\,s^{-1}}$ \citep{Linares_2014, Papitto_2022}, fainter than that of AMXPs in outburst, but brighter than that of quiescent AMXPs and rotation-powered MSPs. Additionally, this state features two X-ray intensity modes, ``high'' and ``low'' modes, with luminosities differing by roughly an order of magnitude, and occasional flaring activity. Mode transitions occur on timescales of about $10~\mathrm{s}$ and each mode persists for periods ranging from $\sim 10~{\rm s}$ to few hours. Based on these characteristic properties of the sub-luminous disk state, many candidate systems have been identified, including several over the past year alone \citep{Pattie_2024, Thongmeearkom_2024, Turchetta_2025, Kyer_2025,Zyuzin_2025}.

PSR~J1023$+$0038, the first discovered and prototype of tMSPs, is characterized by a pulsar spin period of $1.69~{\rm ms}$ and a $0.2 \,{\rm M}_\odot$ main-sequence companion star, orbiting the pulsar in $4.75$ hours \citep{Archibald_2009}. It is observed $\sim 80 \%$ of the time in the high mode \citep{Jaodand_2016}. When in the high mode, X-ray and optical coherent, quasi-simultaneous pulsations are detected \citep{Papitto_2019}. The pulses are similar and characterized by two broad peaks per spin period separated by $\sim 180^\circ$. The optical pulses lag the X-ray ones by $\sim 150 ~{\rm \mu s}$ \citep{Illiano_2023} and are found to be emitted from regions a few tens of kilometers apart. The root-mean-square (rms) pulsed fraction exhibits temporal variability \citep{Miraval_Zanon_2022}. In X-rays the rms amplitudes range from approximately $4.3\%$ to $10.8\%$, whereas in the optical band they span roughly $0.1 \%$ to $0.8 \%$ \citep{Illiano_2023}. Strikingly, the pulsed optical/UV luminosity lies on the same power law as the pulsed X-ray emission \citep{Papitto_2019}, and recent polarimetric observations \citep{Baglio_2025} found that the slope of the polarized emission spectrum matches that of the pulsed emission spectrum. Taken together, these observations suggest that a common mechanism is responsible for the pulsed emission at both optical and X-ray wavelengths, yet its precise origin is still uncertain. 

Since the detection of these puzzling pulsations, numerous studies have tried to unveil their physical origin. Several phenomenological models have been proposed. First, X-ray pulses were attributed to matter channeled onto the magnetic poles by accretion columns \citep{Archibald_2015}. This scenario was challenged by the discovery of corresponding optical pulsations \citep{Ambrosino_2017}, due to their luminosity largely exceeding that expected from such scenario. Such high luminosities also exclude models in which optical photons are produced by accretion-generated X-rays reprocessed at the stellar or disk surface. \citet{Papitto_2015} suggested a propeller model, still relying on accretion-driven X-ray pulsations, for which a significant variation in the spin-down rate between the rotation-powered and disk states would be expected. Conversely, \citet{Ambrosino_2017} proposed a rotation-powered origin of the optical and X-ray pulsations, attributing them to synchro-curvature emission within the pulsar magnetosphere. However, this scenario would imply an exceptionally efficient conversion of spindown power into optical and X-ray emission, far exceeding what is observed in any other pulsar \citep{Papitto_2019}. \citet{Papitto_2019, Veledina_2019} and subsequent works \citep{Campana_2019, Baglio_2025} instead argued that the emission modulated by the pulsar's spin period originates from synchrotron radiation of particles accelerated at the shock front formed between the pulsar wind and the disk. In this scenario, pulsations disappear in the X-ray low mode due to the disk being pushed further out by the pulsar wind. The model put forward by \citet{Veledina_2019} also attributes the pulsations to synchrotron radiation originating from the inner boundary of the accretion disk, lying outside the light cylinder, heated by the pulsar wind. Contrary to the scenario of \citet{Papitto_2019}, the low mode is explained by the disk penetration inside the light cylinder, causing a transition to the propeller regime and suppression of the wind. Recent spin-down rate measurements \citep{Burtovoi_2020, Baglio_2025} show a small ($\sim 5 \%$) decrease when the disk is present, relative to the rotation-powered state of the same source. These results favor the most recent models, as they imply the absence of accretion onto the neutron star surface or of a strong propeller, which would greatly alter this rate, but also a quite close disk, otherwise the spin-down rate would be unchanged.

The aforementioned phenomenological models fundamentally rely on the interaction between the pulsar magnetosphere and the accretion disk. This interaction has been addressed analytically, with notable contributions from \citet{Eksi_2005}, but has been explored primarily through numerical studies, mainly using magnetohydrodynamic simulations. The first insights into disk-magnetosphere coupling were provided by models of magnetized, rotating stars surrounded by an accretion disk, especially in the context of T-Tauri stars \citep{Hayashi_1996, Romanova_2002, Romanova_2003b, Romanova_2008, Zanni_2009, Zanni_2013, Takasao_2018, Romanova_2021}. Accreting neutron stars were first simulated as aligned dipolar rotators \citep{Romanova_2003a, Cikintoglu_2022}, with some studies accounting for general relativistic effects \citep{Parfrey_2017a, Parfrey_2017b,Parfrey_2024}. More recently, these models have been extended to oblique dipolar rotators \citep{Das_2024, Murguia_Berthier_2024} and multipolar magnetic field configurations \citep{Das_2022}. In the past year, a few works have addressed tMSPs using numerical simulations, whether hydrodynamic \citep{Guerra_2024} or magnetohydrodynamic simulations (\citealt{Vurgun_2025}, Mignon-Risse et al., in prep.).\nocite{Mignon_Risse_2025} 

However, these approaches do not account for both the large-scale dynamics of the pulsar wind -- disk interaction and the kinetic processes responsible for particle acceleration and non-thermal radiation. Capturing this combined behavior is essential to explain the non-thermal emission detected from tMSPs, in particular to reproduce their distinctive X-ray and optical signatures. This provides the main motivation for our study and motivates a PIC approach, capable of capturing both the global dynamics and the microphysics of the system.

In the following, we first introduce our numerical model (Section~\ref{section:numerical_model}). In Section~\ref{section:magnetospheric_structure}, we describe the magnetospheric reconfiguration induced by the disk and discuss the associated kinetic effects. Section~\ref{section:sync_rad} presents the observable results, beginning with the luminosity enhancement, followed by the light curves and polarization properties. We then apply our model to the case of PSR J1023$+$0038 in Section~\ref{section:application} and conclude in Section~\ref{section:conclusion}.

\section{PIC numerical model} \label{section:numerical_model}

\subsection{3D initial setup} \label{subsection:initial_setup}

We resort to the 3D spherical, flat spacetime version of the {\tt Zeltron} code. The grid, of spherical coordinates ($r$, $\theta$, $\phi$), is made of $1024\times128\times256$ cells in the respective directions. Uniform angular spacing is adopted in $\theta$ and $\phi$, while the radial direction is sampled logarithmically to preserve the cell aspect ratio with radius. This choice is well suited to pulsar magnetosphere simulations, where particle densities and field strengths decline with radius, causing both the plasma skin depth and the Larmor radius to grow with radius. The neutron star surface ($r=r_\star$) defines the inner boundary of the domain, which extends up to $r_{\rm max}=5.6\, R_{\rm LC}$, where $R_{\rm LC}= c/\Omega = 5\,r_\star$ is the light-cylinder radius and $\Omega$ is the star's angular velocity. The outer boundary is designed to capture the immediate wind--disk interaction region, but it is not intended to model radiative evolution over the larger scales that may contribute to inter-band lags and other non-local observables. We include an absorbing layer at $r_{\rm absorb}=0.9 r_{\rm max}$ for fields and particles, following the approach of \citet{Cerutti_2015}, to approximate open-boundary conditions. Given the parameters adopted in these simulations, the fast magnetosonic point lies close to the outer edge of the computational domain, which justifies treating the wind as freely expanding to infinity beyond that radius (see \citealt{Richard_Romei_2024} for a comparison of the sub- and super-magnetosonic regimes). Particles impacting the star surface are also absorbed, while elastic reflection is applied at the $\theta-$boundaries, and periodic boundary conditions are imposed in the $\phi-$direction. For the fields, axial symmetry is enforced along $\theta$, and periodic boundary conditions are implemented in the $\phi-$direction. 

The simulation begins with a dipolar magnetic field ($\mathbf{B}$) anchored to the star surface, in vacuum. The rotation axis ($\mathbf{\Omega}/\Omega$) is aligned with the $\theta=0$ direction and the magnetic moment ($\boldsymbol{\mu}$) is inclined by an angle $\chi$ in the $\phi=0$ plane relative to the rotation axis. 
The magnetic field is expressed as:
\begin{align}
    \mathbf{B} (\mathbf{r}) &=\frac{3 \left(\boldsymbol{\mu}\cdot\mathbf{r}\right)\mathbf{r}}{r^5}-\frac{\boldsymbol{\mu}}{r^3} \\
	& \quad = \frac{\mu}{r^3} \left(2\sin\chi\sin\theta\cos\left(\Omega t-\phi\right)+2\cos\chi\cos\theta\right)\mathbf{u}_r \notag\\
	& \quad + \frac{\mu}{r^3} \left(-\sin\chi\cos\theta\cos\left(\Omega t - \phi\right)+\cos\chi\sin\theta\right)\mathbf{u}_\theta \notag \\
	& \quad - \frac{\mu}{r^3}\sin\chi\sin\left(\Omega t -\phi\right)\mathbf{u}_\phi\,, \notag
\end{align}
where the norm of the magnetic moment is given by $\mu=B_\star r_\star^{3}$, with $B_\star$ denoting the surface polar magnetic field.
The perfect conductor condition imposed at the star surface induces, due to the Lorentz transformation, an electric field:
\begin{equation}\label{eq:corotation_efield}
\mathbf{E}=-\frac{(\mathbf{\Omega}\times \mathbf{r}_\star)\times\mathbf{B}}{c} \,.
\end{equation}
This constraint causes the magnetic field lines to start rotating at $t=0$. 

Initially, the simulation domain is devoid of plasma. Plasma is injected via two complementary mechanisms. First, pairs are supplied uniformly across the entire star surface, at a rate of one macroparticle per cell, with a probability of injection set at $20\%$. Each simulated macroparticle represents a large number of physical particles, characterized by a weight ($w_p$), all having the same charge-to-mass ratio and thus following identical trajectories in phase space. To ensure the formation of electric gaps able to accelerate particles, plasma injection takes place only in regions where the plasma density is lower than the Goldreich-Julian density ($n_{\rm GJ}=\Omega B_\star/2\pi e c$; \citealt{Goldreich_1969}). In those regions, a charge density given by $10\%$ of the difference between the radial electric field and the corotation radial electric field (Eq.~\ref{eq:corotation_efield}), both computed at the star surface, is released at each timestep, in corotation with the star. Secondly, pairs are also supplied wherever the Lorentz factor of the primary particle exceeds a threshold, defined by $\gamma_{\rm th}=0.005 \gamma_{\rm pc}$, where $\gamma_{\rm pc}=e\mu\Omega^{4}/m_{\rm e}c^{4}$ denotes the maximum Lorentz factor reachable by a particle accelerated by the polar cap potential drop and $e$, $m_{\rm e}$ are respectively the charge and mass of the electron. Pairs are injected at the location of the primary particle and along its direction of motion, without intermediate photon emission, assuming a negligible photon mean free path. Newly-created pairs all display the same energy, given by $0.1\gamma_{\rm th}$, and the same amount of energy is removed from the parent particle. Similarly to the first injection mechanism, plasma injection is inhibited wherever the charge density exceeds a threshold density, fixed at $10\, n_{\rm GJ}$. Ions are neglected in this setup, due to their expected little impact on the magnetospheric dynamics and on the radiation. 

The large-scale difference between plasma kinetic scales and macroscopic magnetospheric scales requires us to rescale the simulations. Indeed, we obtain a ratio between the skin depth at the star surface and the star radius of $d_{\rm e}^{\star}/r_\star \sim 7 \times 10^{-3}$, implying a simulation rescaling of approximately three orders of magnitude compared to actual spatial scales. The plasma skin depth is resolved by roughly 2 cells ($\Delta r$) at the stellar surface, about 18 cells at $R_{\rm LC}$, and gradually saturates to approximately 20 cells at larger radii. At the light-cylinder radius, the Larmor radius is resolved by about 13 cells within the current sheet and 1 cell in the wind. Its resolution in the wind increases with radius, reaching roughly 5 cells at $r_{\rm max}$. Table~\ref{tab:init_params} summarizes the main simulation parameters.

 \begin{table}[ht!]
	\caption{Simulation parameters.}
	\centering
	\begin{tabular}{lc}
		\hline
		\hline
		\noalign{\smallskip}
		Parameter & \quad Value \\
		\noalign{\smallskip}
		\hline
		\noalign{\smallskip}
		Number of cells & \qquad $1024\,(r) \times 128\,(\theta) \times 256\,(\phi)$  \\
		Inner boundary   & \qquad  $r_\star$ \\
		$R_{\rm LC}$  & \qquad 5 $r_\star$\\
		$r_{\rm max}$  & \qquad 5.6 $R_{\rm LC}$ \\
        $d^{\star}_{\rm e}/r_\star$ & \qquad  $7 \times 10^{-3}$\\
		$(d_{\rm e}/\Delta r)_{\rm LC}$ & \quad 18   \\
		$P_{\rm spin} / \Delta t$ & \qquad  $9.5 \times 10^{4}$ \\
		Plasma composition & \qquad electrons + positrons \\
		Injection model & \qquad from star surface + pair creation \\
        $\chi (^\circ)$ & \qquad 15, 30, 60, 75 \\
        $R$ & \qquad  $3.572\,R_{\rm LC}$ \\
        $h$ & \qquad  $0.3\,R$ \\
		\noalign{\smallskip}
		\hline
	\end{tabular}
	\label{tab:init_params}
\end{table}

\subsection{Disk implementation}\label{subsection:disk_implementation}

\begin{figure}[bthp]
\centering 
\includegraphics[width=\columnwidth, keepaspectratio]{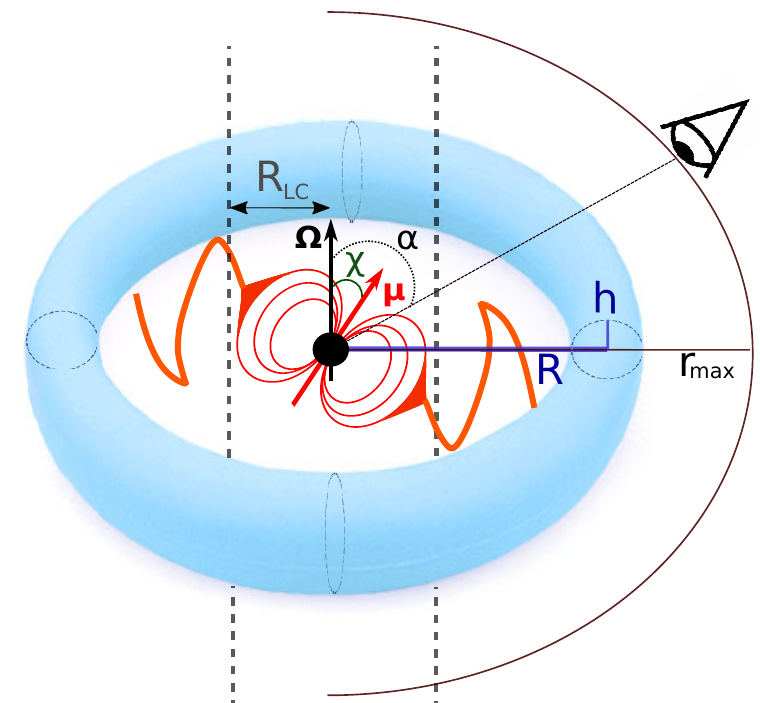}
\caption{Numerical setup. The pulsar, of angular velocity $\mathbf{\Omega}$ and light-cylinder radius $R_{\rm LC}$ is represented by the black sphere. The magnetic moment $\mathbf{\mu}$ is inclined by an angle $\chi$ with respect to the rotation axis. The torus aspect ratio $h/R$ is of $0.3$. $\alpha$ denotes the observer's viewing angle and $r_{\rm max}$ the outer boundary of the simulation box.}\label{fig:sketch_setup}
\end{figure}

The disk is modeled as a perfectly conducting torus (see Figure~\ref{fig:sketch_setup}). The disk is positioned in the equatorial plane, with its inner radius at $2.5\,R_{\rm LC}$, which is in agreement with the radii expected in the case of tMSPs \citep{Papitto_2019}. Moreover, such a distance represents a good compromise, allowing the proper formation of the current sheet before the disk but also keeping sufficient space beyond the disk to observe the rearrangement of the magnetosphere. The aspect ratio of the disk is not well constrained observationally \citep{Papitto_2019}. We set $h/R=0.3$, where $h$ denotes the minor radius of the torus and $R=3.572\, R_{\rm LC}$ its major radius (see Figure~\ref{fig:sketch_setup}), consistently with previous models \citep{Parfrey_2017b, Veledina_2019}. The magnetic field of the disk is neglected: $\mathbf{B}_{\rm torus}=\mathbf{0}$. We also neglect its orbital motion, the Keplerian period being much longer than the pulsar's spin period; hence $\mathbf{E}_{\rm torus}=\mathbf{0}$. The disk absorbs any particle or photon hitting its surface. Hence, the disk is not evolved as a plasma flow, but instead serves as a boundary condition for the pulsar wind. The setup is designed to isolate the electrodynamic effects of a nearby conducting disk on the pulsar wind.

\subsection{Temporal evolution scheme}
The self-consistent evolution of particles and fields relies on a computational loop completed at each timestep. Particles evolve according to the Abraham-Lorentz-Dirac equation, which contains a radiation-reaction term to account for radiative losses. An area-weighting scheme enables to deposit charges and currents on the grid, before evolving the fields. Electromagnetic fields evolve on a grid according to the Maxwell-Faraday and Maxwell-Ampère equations through the Yee algorithm \citep{Yee_1966}. The Poisson equation is solved with the Gauss-Seidel method every $25$ timesteps to ensure charge conservation. 
The timestep ($\Delta t$) is set to half the Courant-Friedrich-Lewy stability limit and the spin period amounts to $P_{\rm spin} \sim 9.5\times 10^{4} \Delta t$.
General relativistic effects are neglected, as our focus is on the pulsar wind rather than the immediate vicinity of the star surface.

\subsection{Synchrotron-curvature modeling}

In the studied systems, the cooling of particles is predominantly due to curvature and synchrotron radiation. We therefore restrict this study to the modeling of synchro-curvature emission. To model such radiation, we consider that each macroparticle emits a macrophoton, which corresponding number of physical photons is determined by the macroparticle's weight ($w_p$). Each macrophoton radiates the following power:
\begin{equation}\label{eq:prad}
    \mathcal{P}_{\rm rad}=\frac{2}{3}w_p r_{\rm e}^{2}c\gamma^{2}\Tilde{B}^{2}_\perp \,,
\end{equation}
where $r_{\rm e}$ is the classical radius of the electron, $\gamma$ is the particle's Lorentz factor and  $\Tilde{B}^{2}_\perp$ is the effective perpendicular magnetic field, defined as \citep{Cerutti_2016}
\begin{equation}\label{eq:bperp}
    \Tilde{B}_\perp=\sqrt{(\mathbf{E}+\boldsymbol{\beta}\times\mathbf{B})^{2} - (\boldsymbol{\beta} \cdot \mathbf{E})^{2}} \,.
\end{equation}
Contrary to the classical expression used in the case of pure magnetic fields $B_\perp=B \sin \alpha$, with $\alpha$ being the angle between the magnetic field and the particle's direction of motion, $\Tilde{B}^{2}_\perp$ takes into account both the electric and magnetic fields, which are of comparable strength in the pulsar wind. Photons are emitted along the particle's direction of motion, due to the strong relativistic beaming, and propagate along straight trajectories at the speed of light. They do not interact with each other nor with the magnetic field. Photons are eclipsed by the star, and for the purpose of computing photon trajectories, the outer side of the torus is extended to the outer boundary of the domain at fixed $h$, ensuring that photons intersecting either the torus or this extension are eclipsed. Non-eclipsed photons are registered on a plane perpendicular to the observer's line of sight and tangent to the outer edge of the simulation domain, serving as a proxy for a screen located at infinity.

\section{Magnetospheric structure}\label{section:magnetospheric_structure}

\subsection{Dependence on magnetic obliquity}
The latitudinal opening angle of the current sheet is directly determined by the magnetic obliquity of the pulsar ($\chi$). Since the current sheet represents the densest region of the magnetosphere beyond the light cylinder, the fraction of the sheet intercepted by the disk is expected to influence both the global restructuring of the magnetosphere and the associated electromagnetic emission. Therefore, keeping the disk aspect ratio fixed at $h/R=0.3$, corresponding to a disk half-opening angle of 16.7°, we explore the parameter space by performing simulations for magnetic obliquities $\chi=15^\circ$, $30^\circ$, $60^\circ$, and $75^\circ$. For comparison, we also run the four corresponding reference simulations for isolated pulsars. Simulations were run for $5\,P_{\rm spin}$ and converged after $1.1\,P_{\rm spin}$. This was evidenced by a transient in the outgoing radial Poynting flux, generated at the stellar surface, leaving the computational domain.

\subsection{Density}

\begin{figure*}[bthp]
\centering 
\includegraphics[width=\columnwidth, keepaspectratio]{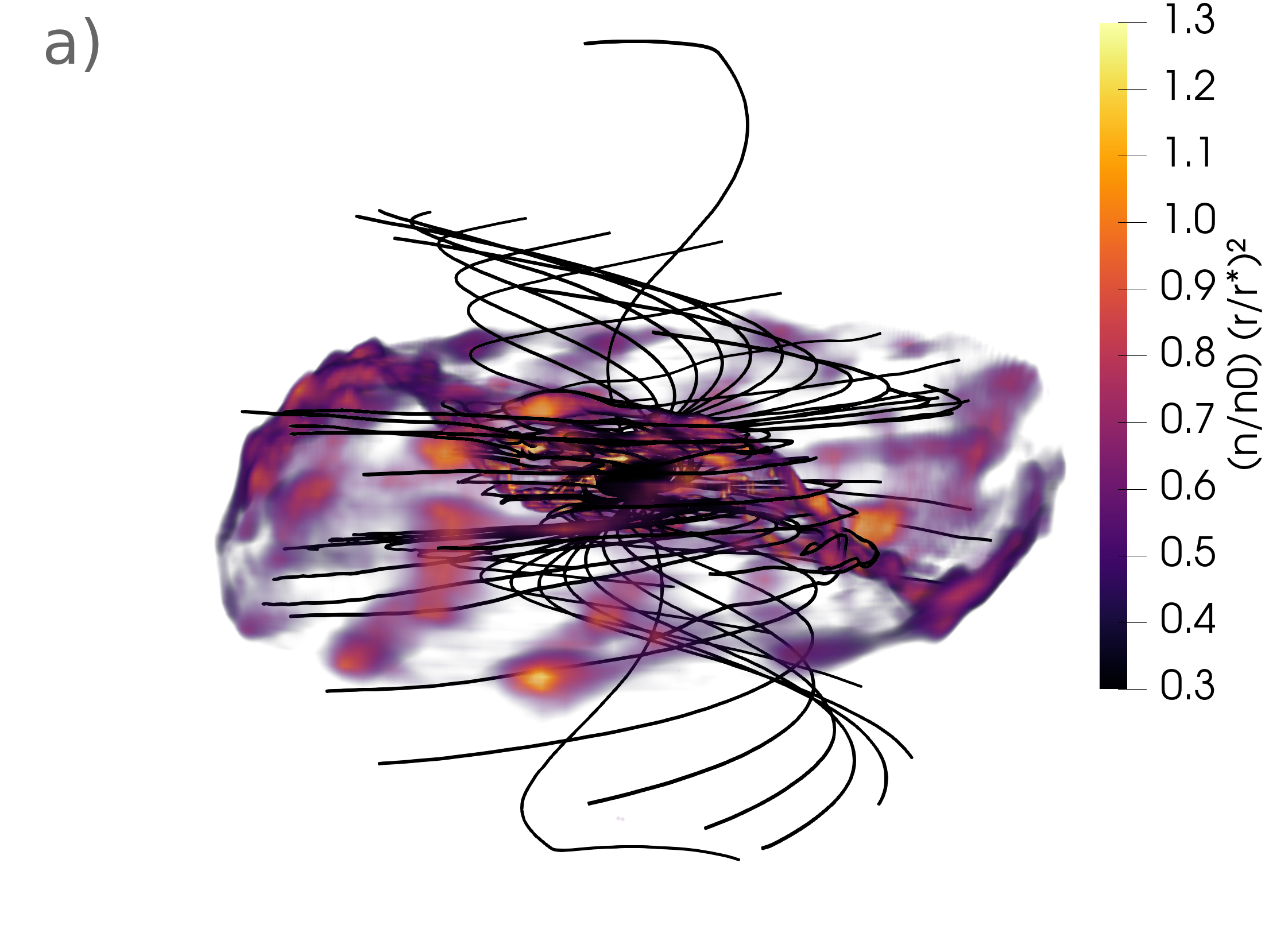}
\includegraphics[width=\columnwidth, keepaspectratio]{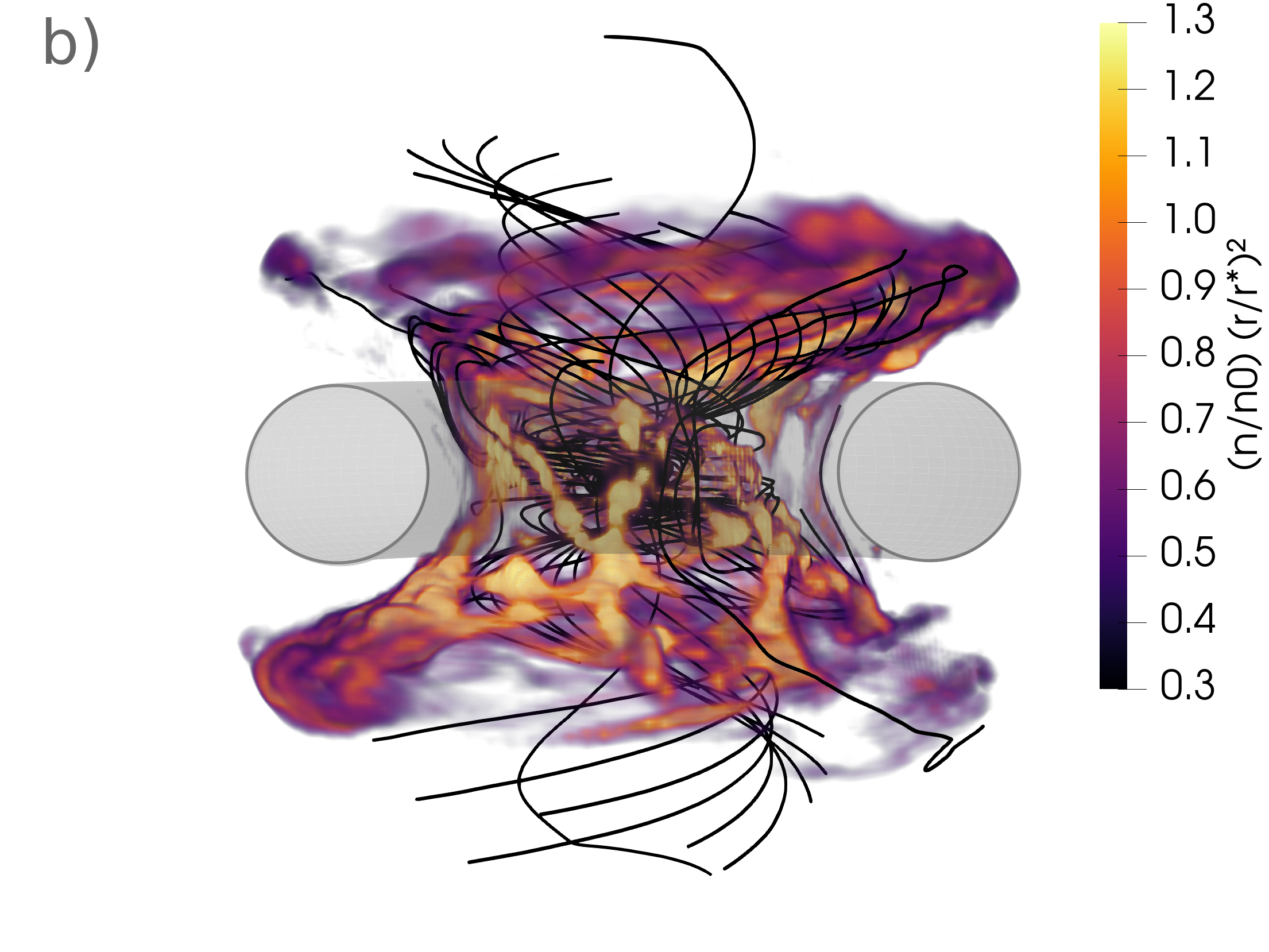}
\includegraphics[width=\columnwidth, keepaspectratio]{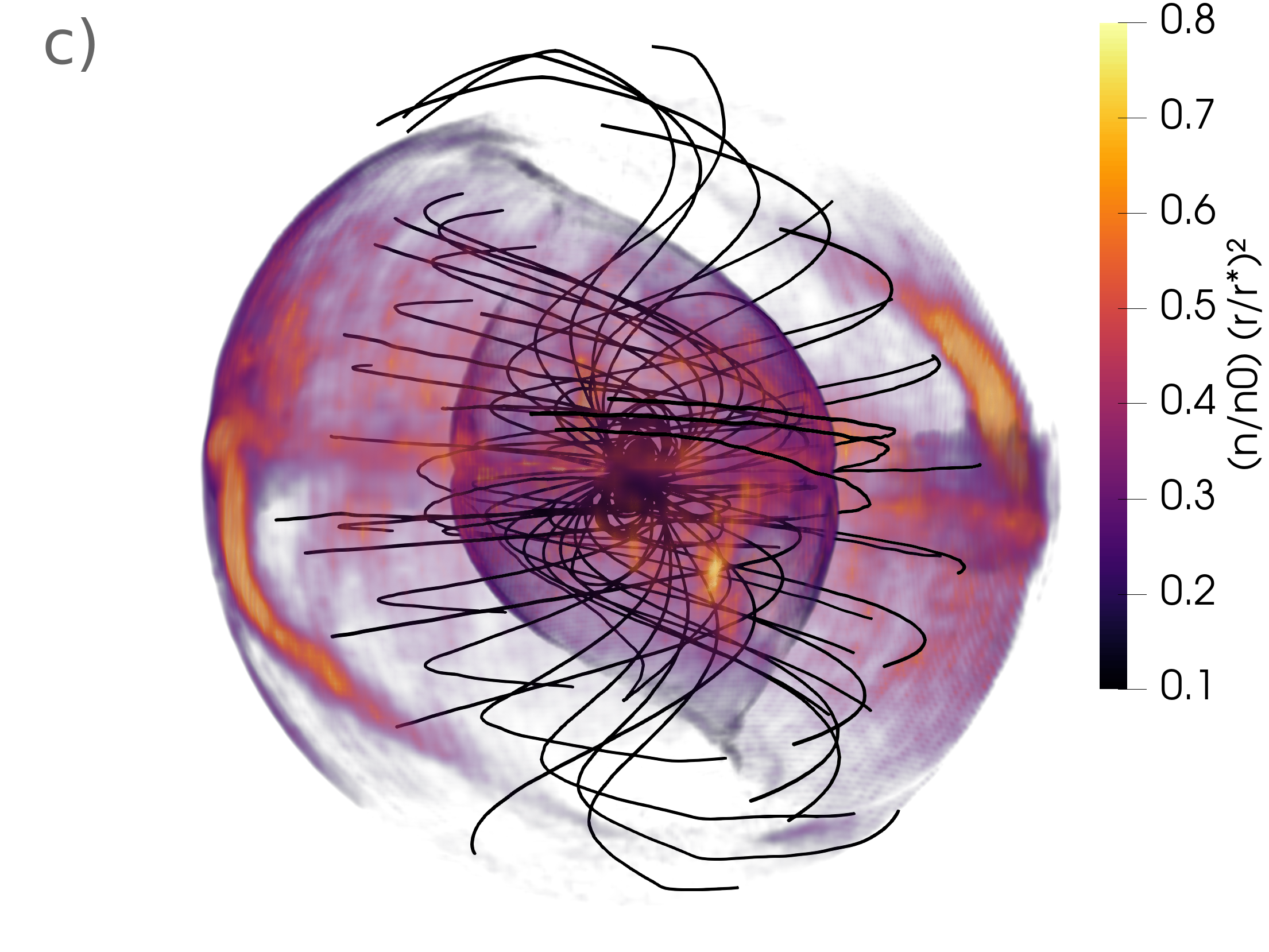}
\includegraphics[width=\columnwidth, keepaspectratio]{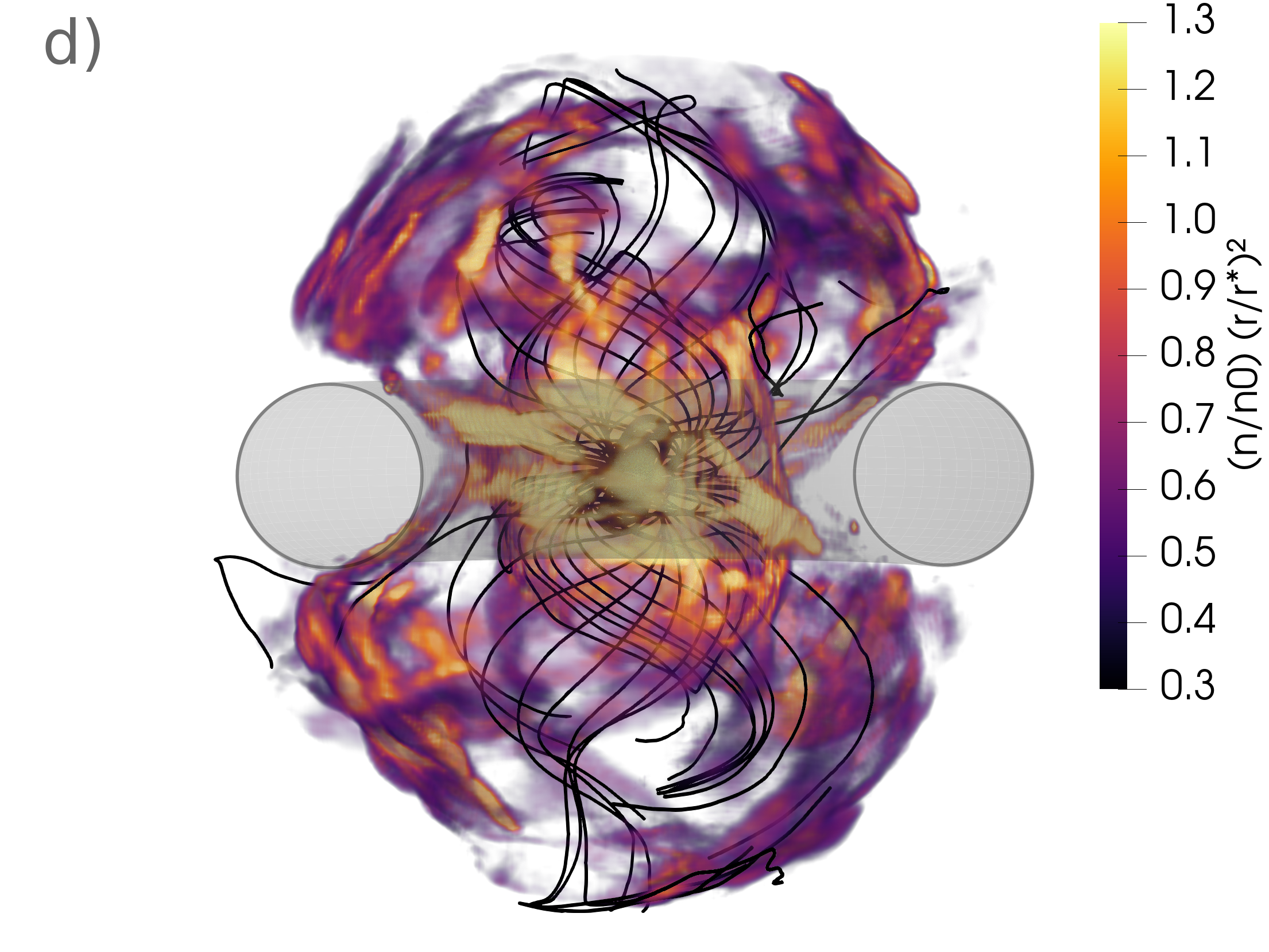}
\caption{3D rendering of the particle density $(n/n_0)(r/r_\star)^{2}$ in pulsar magnetospheres at $t=2.5\,P_{\rm spin}$. The normalization factor is defined as $n_0=\kappa \,n_{\rm GJ}$, where $\kappa=10$ represents the plasma multiplicity and $n_{\rm GJ}$ corresponds to the Goldreich-Julian density computed at the stellar magnetic poles. The central black sphere represents the neutron star, and black solid lines trace the magnetic field lines. \textit{Panels (a) and (c):} isolated pulsar magnetospheres for $\chi=15^\circ$ and $\chi=75^\circ$, respectively. \textit{Panels (b) and (d)}: reconfiguration of the pulsar magnetosphere in the presence of the torus (grey shaded region), for $\chi=15^\circ$ and $\chi=75^\circ$, respectively. }\label{fig:3D_structure}
\end{figure*}

\begin{figure*}[bthp]
\centering 
\includegraphics[width=\textwidth, keepaspectratio]{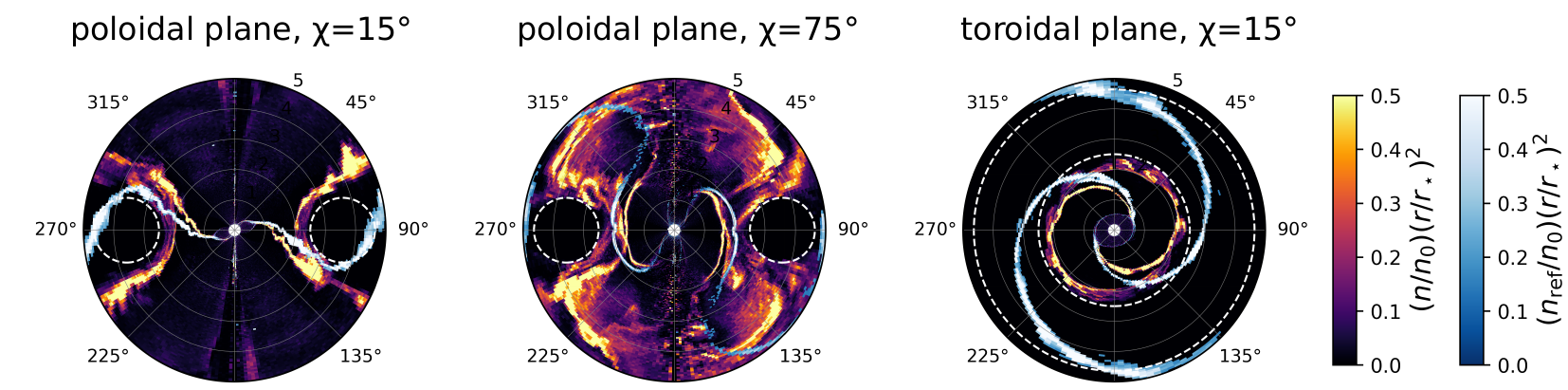}
\caption{2D snapshots of the plasma density $(n/n_0)(r/r_\star)^{2}$, after $2.5$ pulsar spin periods. The left and middle panels show poloidal cuts ($\phi=0^\circ$) for a magnetic obliquity of $\chi=15^\circ$ and $\chi=75^\circ$ respectively, while the right panel shows a toroidal cut ($\theta=0^\circ$) for $\chi=15^\circ$, which is representative of the other obliquity cases. We superimpose the current sheet density obtained in the isolated case ($n_{\rm ref}$; blue color scale) onto the plasma density ($n$; warm color scale) resulting from the interaction case. Radii are given in units of $R_{\rm LC}$.}\label{fig:2D_density}
\end{figure*}

The 3D density structure of the magnetosphere is presented in Figure~\ref{fig:3D_structure}. Panels (a) and (c) show the isolated pulsar cases for $\chi=15^\circ$ and $\chi=75^\circ$, respectively. The current sheet forms a 3D Archimedean spiral of wavelength $2\pi R_{\rm LC}$ \citep{Coroniti_1990, Bogovalov_1999}. As expected, increasing the magnetic obliquity enhances the latitudinal expansion of the current sheet, which extends over an angle $2\chi$ in the $\theta$-direction. We observe entangled plasma flux ropes produced by the fragmentation of the current sheet through the tearing instability. The addition of the disk around the torus significantly alters the magnetosphere (Figure~\ref{fig:3D_structure}, panels b and d), as it forces the rearrangement of both plasma and magnetic field lines around the obstacle. The plasma density increases markedly, owing to the partial confinement of the plasma between the star and the inner edge of the disk. As a result, the magnetosphere develops vertically. Some plasma manages to circumvent the disk by flowing along field lines, thereby reaching larger radii. The confinement of the plasma and the increased vertical extent of the plasma distribution around the torus imply a geometrically thicker emitting region than in the isolated case. This is expected to enhance the quasi-continuous component of the emission, broaden the pulse profiles by keeping portions of the thicker current sheet visible over a larger fraction of the spin cycle, and reduce the net polarization through the superposition of emission from a wider range of magnetic-field orientations. For sufficiently large magnetic obliquities such that the magnetic poles are oriented toward the disk surface (with the critical angle determined by the disk aspect ratio), some open field lines anchored at the polar caps are periodically driven to close and subsequently reopen. This cyclic process modulates the plasma dynamics and the associated energy release, producing alternating phases of retention and ejection. For visualization purposes, 2D poloidal maps of the plasma density, in which the isolated and pulsar-disk interaction cases are superimposed, are shown in Figure~\ref{fig:2D_density}, for both $\chi=15^\circ$ and $\chi=75^\circ$. In the least inclined case (Fig.~\ref{fig:2D_density}, left panel), the original shape of the current sheet closely matches the inner edge of the torus, causing the plasma to be channeled along the inner boundary of the disk with only minor perturbations.
For highly inclined configurations (Fig.~\ref{fig:2D_density}, middle panel), the current sheet intersects the torus more vertically, strengthening perturbations of the magnetic field lines and plasma flow, and thereby yielding a more turbulent reconfiguration of the magnetosphere. For all magnetic obliquities, the disk induces a tighter winding of the two spiral arms of the current sheet around the pulsar, leading to their merging at the inner edge of the disk, as illustrated by the 2D equatorial map shown in Figure~\ref{fig:2D_density}, right panel.

\subsection{Magnetic field and kinetic effects}

\begin{figure*}[bthp]
\centering 
\includegraphics[width=\textwidth, keepaspectratio]{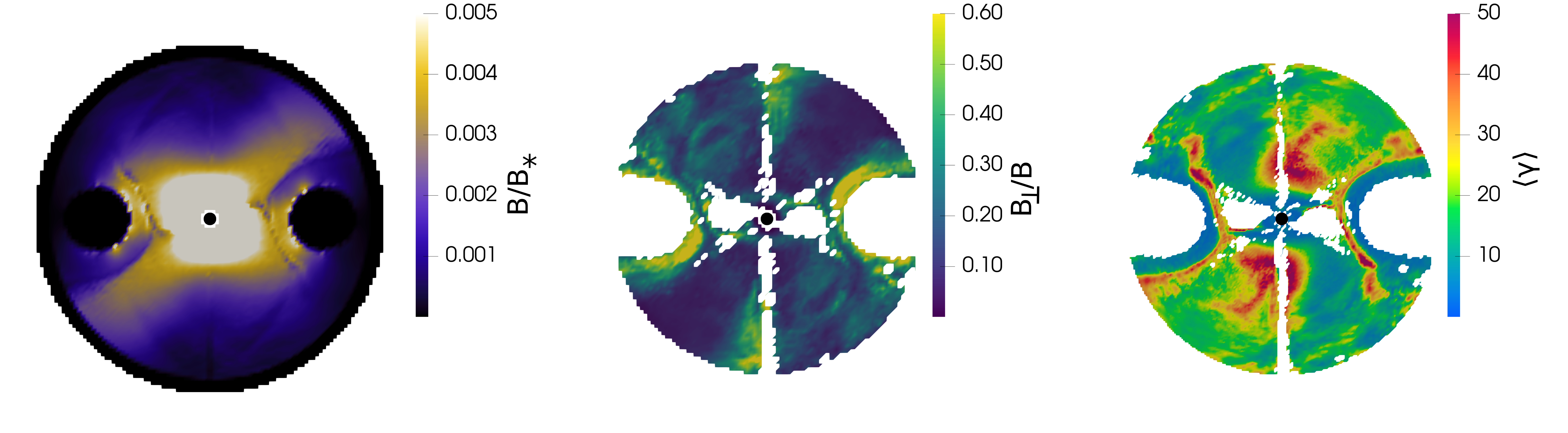}
\caption{2D poloidal maps of the pulsar-disk interaction case for $\chi=15^\circ$, at $1.4\,P_{\rm spin}$. \textit{Left panel:} magnetic field strength ($B/B_\star$). \textit{Middle panel:} plasma isotropization ($B_\perp/B$). \textit{Right panel:} particle acceleration ($\langle \gamma \rangle$). The central black sphere represents the neutron star.}\label{fig:diagnostics}
\end{figure*}

The change in the magnetospheric structure leads to a significant increase in the magnetic field strength, most prominently at the inner edge of the torus, but extending throughout the entire simulation domain (see Fig.~\ref{fig:diagnostics}, left panel). Plasma isotropization is also enhanced in the presence of the disk, as shown in Figure~\ref{fig:diagnostics} (middle panel). In isolated pulsars, the plasma globally follows the magnetic field lines, except when trapped in the current sheet, where it is isotropized by acceleration processes. The introduction of the disk, however, alters particle trajectories, thereby amplifying pitch angles by at least a factor of two near the disk edges. Increasing the magnetic obliquity further perturbs the current sheet, leading to an even stronger plasma isotropization. Another important effect of the disk is the acceleration of particles at its inner edge (see Fig.~\ref{fig:diagnostics}, right panel). Indeed, the torus surface causes the bending and compression of magnetic field lines, which favors the dissipation of magnetic energy into particles, via magnetic reconnection. The enhancement of particle acceleration at the polar caps is likely related to the modification of the plasma boundary conditions induced by the presence of the torus. In this configuration, charge conservation at the pulsar surface appears to require an increased charge density in the current sheet.

\section{Synchrotron radiation}\label{section:sync_rad}

\subsection{Increase of the synchrotron power}\label{section:incr_sync}

\begin{figure*}[bthp]
\centering 
\includegraphics[width=\textwidth, keepaspectratio]{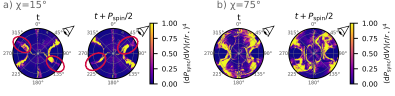}
\caption{Time evolution of 2D poloidal synchrotron maps $\left({\rm d}P_{\rm sync}/{\rm d}V\right)(r/r_\star)^{4}$, illustrating the effect of the disk for $\chi=15^\circ$ (panel a) and $\chi=75^\circ$ (panel b), shown at times $t=2.5\,P_{\rm spin}$ and $t+P_{\rm spin}/2$. Magenta circles indicate emission minima, and the eye sketch marks the observer's viewing angle. Radii are given in units of $R_{\rm LC}$.}\label{fig:psyn_evol}
\end{figure*}

The rearrangement of the interacting magnetosphere described in Section~\ref{section:magnetospheric_structure}, together with the associated magnetic and kinetic effects, implies (see Eq.~\ref{eq:prad}) a substantial increase in the synchrotron emission in the presence of the disk. The disk not only enhances the synchrotron radiation but also modifies the emission region, which now corresponds to the geometry of the disrupted current sheet. As a result, a marked structural difference arises between low- and high-obliquity cases, as illustrated in Figure~\ref{fig:psyn_evol}. For low-obliquity cases, each arm of the current sheet sweeps from $\pi/2+\chi$ to $\pi/2-\chi$, crossing the observer's angle once per rotation period. The radially escaping emission thus peaks at time $t$ and reaches a minimum at $t+P_{\rm spin}/2$ (see magenta circles on Fig.~\ref{fig:psyn_evol}, panel a), with the same pattern shifted by half a spin period on the opposite hemisphere. Overall, the emission is modulated over a pulsar rotation, yielding a single intensity peak per spin period. Conversely, for high-obliquity cases (see Fig.~\ref{fig:psyn_evol}, panel b), the nearly vertical orientation of the current sheet as it reaches the inner edge of the disk causes it to break apart around the disk in an approximately symmetric manner. Consequently, each time an arm of the current sheet reaches the disk, a burst of light is emitted, regardless of the hemisphere. We therefore expect two intensity peaks per spin period, for a given observer's angle.

\subsection{Light curves}
\begin{figure*}[bthp]
\centering 
\includegraphics[width=0.8\textwidth, keepaspectratio]{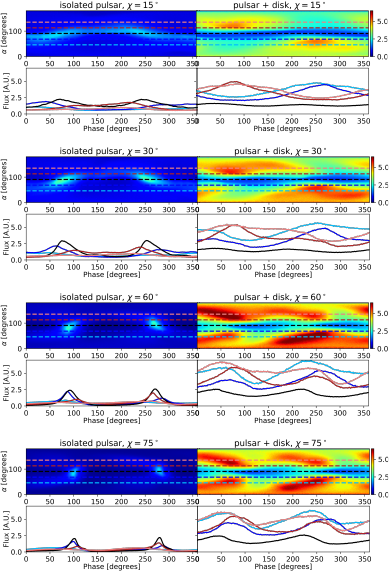}
\caption{Skymaps of the synchrotron radiation as a function of the viewing angle ($\alpha$) and the pulsar rotation phase, for all simulations. Below each skymap, we show the corresponding light curves computed for five different observer angles, $\alpha= 45^\circ$, $67.5^\circ$, $90^\circ$, $112.5^\circ$ and $135^\circ$, with dashed lines on the skymaps indicating the respective colors.  The synchrotron flux is normalized to the time-averaged flux of the $\chi=15^\circ$ reference simulation. Data are averaged over $2\,P_{\rm spin}$ after convergence.}\label{fig:skymaps}
\end{figure*}

As suggested by Section~\ref{section:incr_sync}, the introduction of the disk in the pulsar magnetosphere has dramatic effects on the emitted light curves. The most striking observational signature emerging from the simulations is the replacement of the narrow current-sheet pulses seen in isolated pulsars by broader, brighter, disk-induced pulses, whose number and morphology depend primarily on how the current sheet interacts with the disk. Figure~\ref{fig:skymaps} presents the time-averaged skymaps and light curves computed for both the reference isolated simulations (left column) and the pulsar-disk interaction simulations (right column), for each magnetic obliquity. For the reference simulations, we recover well-established results (e.g., \citealt{Cerutti_2016}): the light curves display two narrow intensity peaks originating from the current sheet. The width and phase separation of these peaks depend on the geometry of the current sheet as intercepted by the observer's line of sight, and therefore on the pulsar's magnetic obliquity. The light curves obtained in the interaction case differ markedly from the canonical ones. The intensity increases substantially throughout the entire rotation period, independently of both the magnetic obliquity and the observer's viewing angle. The intensity peaks are considerably broader than in the isolated pulsar case, regardless of the magnetic obliquity. When two peaks are present, they exhibit slight differences in amplitude and are separated by half a spin period. We observe an increase in intensity with magnetic obliquity, although the $\chi=60^\circ$ simulation appears to be the brightest. In agreement with Figure~\ref{fig:psyn_evol}, the low-obliquity case ($\chi = 15^\circ$) exhibits a single peak per spin period for a given viewing angle, with the emission from the two hemispheres separated by half a spin period in phase. In contrast, the high-obliquity case ($\chi=75^\circ$) exhibits two peaks per period that occur at the same rotational phase in both hemispheres. The intermediate magnetic obliquities (i.e., $\chi=30^\circ$ and $\chi=60^\circ$) show a progressive transition from one regime to the other.

\begin{figure}[bthp]
\centering 
\includegraphics[width=\columnwidth, keepaspectratio]{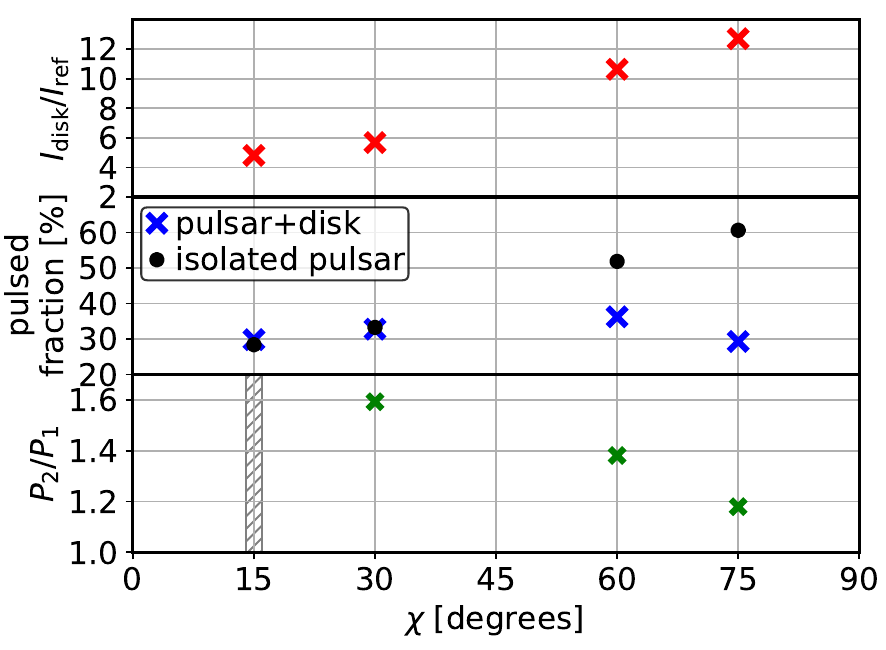}
\caption{Light curves parameters as a function of the magnetic obliquity. \textit{Top panel:} ratio of the total luminosity in the interaction case ($I_{\rm disk}$) to that in the isolated case ($I_{\rm ref}$). \textit{Middle panel:} pulsed fraction obtained by subtracting the continuous component to the total intensity, and normalizing by the total intensity. \textit{Bottom panel:} ratio of the peak amplitudes ($P_2/P_1$). The dashed column indicates the absence of a second peak. Data are averaged over $2\,P_{\rm spin}$ after convergence.}\label{fig:obs_params}
\end{figure}

Figure~\ref{fig:obs_params} summarizes the dependence of these properties on magnetic obliquity. Firstly, the total luminosity emitted in the interaction case is enhanced by a factor of $\sim 5$ to $\sim 12.5$ relative to the corresponding reference intensity, as $\chi$ increases. Secondly, the pulsed fraction, computed by subtracting the continuous flux  (i.e., the phase minimum) from the total flux and normalizing to the total flux, remains significantly lower than in the reference simulations. This is due to the presence of a strong continuous component. While the pulsed fraction increases with magnetic obliquity in isolated pulsars, it remains roughly constant at $\sim 30\%$ in the interaction case. The peak amplitude ratios $P_2/P_1$ decrease from 1.6 to 1.2 as $\chi$ increases from $30^\circ$ to $75^\circ$.

\subsection{Polarization signatures}
The modification of the magnetic field topology caused by the disk affects the polarization signatures of the system. Due to the ultra-relativistic nature of particles, the circular polarization component is neglected. The polarization degree is then given by
\begin{equation}
    \Pi=\frac{\sqrt{Q^{2}+U^{2}}}{I}\,,
\end{equation}
where $I$, $Q$ and $U$ are the Stokes parameters.
The polarization angle (PA), defined between the rotation axis and the direction of the effective perpendicular magnetic field ($\Tilde{B}_\perp$; Eq.~\ref{eq:bperp}), reads:
\begin{equation}
    \tan (2PA)=\frac{U}{Q}\,.
\end{equation}
A detailed description of their computation is provided in \citet{Cerutti_2016bis}.

\begin{figure}[bthp]
\centering 
\includegraphics[width=\columnwidth, keepaspectratio]{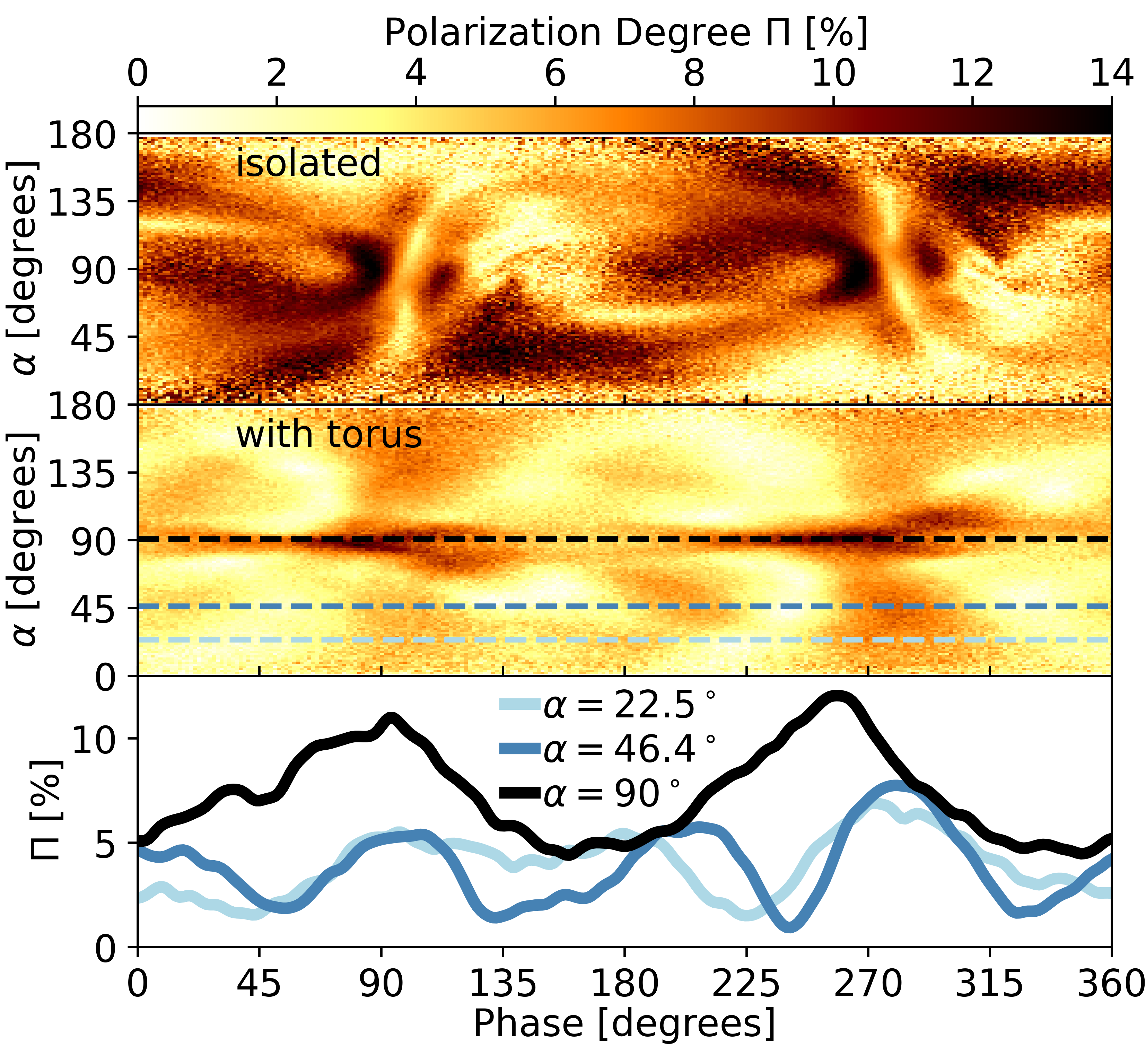}
\caption{Polarization degree ($\Pi$) as a function of the observer's viewing angle ($\alpha$) and the rotation phase, for $\chi=75^\circ$. \textit{Top panel:} isolated pulsar magnetosphere. \textit{Middle panel:} pulsar-disk interaction case. The dashed lines indicate specific viewing angles, with the corresponding curves shown in the bottom panel. \textit{Bottom panel:} polarization degree for three specific viewing angles ($\alpha=22.5^\circ$, $46.4^\circ$ and $90^\circ$). Data are averaged over $2\,P_{\rm spin}$ after convergence.}\label{fig:DP}
\end{figure}

Figure~\ref{fig:DP} shows the polarization degree obtained for the $\chi=75^\circ$ simulations, both for the isolated case and in the presence of the torus, considering all observer's viewing angles and rotation phases. Although we observe some variations with magnetic obliquity, some features are common to all magnetic obliquities. The isolated case (Fig.~\ref{fig:DP}, top panel) is characterized by a polarization degree reaching up to $20\%$, and a pronounced drop in polarization within the current sheet, due to the magnetic field cancellation at reconnection X-points, and the strong isotropization of accelerated particles. The introduction of the torus (Fig.~\ref{fig:DP}, middle panel) induces a depolarization of the magnetosphere, as it causes the disruption of the current sheet, the accumulation of plasma before its inner edge, and the isotropization of particles. Accordingly, a correlation emerges between the flux intensity peaks (Fig.~\ref{fig:skymaps}, right bottom panel) and the local minima of the polarization degree.

\begin{figure}[bthp]
\centering 
\includegraphics[width=\columnwidth, keepaspectratio]{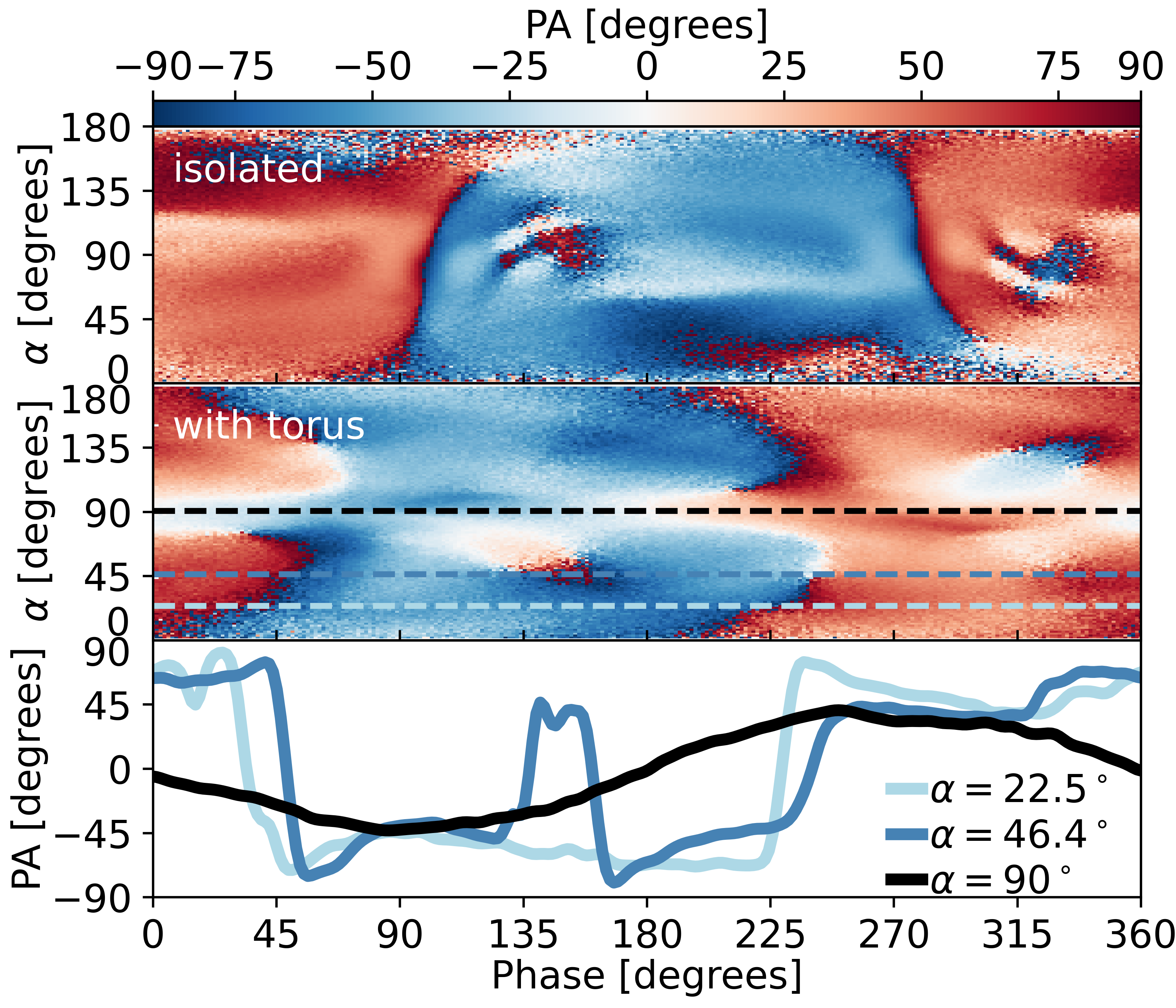}
\caption{Polarization angle (PA) as a function of the observer's viewing angle ($\alpha$) and the rotation phase, for $\chi=75^\circ$. \textit{Top panel:} isolated pulsar magnetosphere. \textit{Middle panel:} pulsar-disk interaction case. The dashed lines indicate specific viewing angles, with the corresponding curves shown in the bottom panel. \textit{Bottom panel:} polarization angle for three specific viewing angles ($\alpha=22.5^\circ$, $46.4^\circ$ and $90^\circ$). Data are averaged over $2\,P_{\rm spin}$ after convergence.}\label{fig:AP}
\end{figure}

Similarly, Figure~\ref{fig:AP} is devoted to the polarization angle. For isolated pulsars (Fig.~\ref{fig:AP}, top panel), the dominant feature is the sharp polarization angle swing occurring across the current sheet, caused by the reversal of the magnetic field across it. On either side of the current sheet, the magnetic polarity remains nearly uniform. Adding the disk (Fig.~\ref{fig:AP}, middle panel) does not significantly affect the rotation of the polarization angle. Indeed, it rotates approximately from $-70^\circ$ to $70^\circ$ across the current sheet, and remains nearly uniform on either side of the current sheet. However, the equatorial regions are rather blurred, with lower polarization angles, due to the eclipse. The disruption of the current sheet causes the local mixing of different polarities, thereby leading to some additional noise. These results are consistent across all magnetic obliquities. 

\section{Application to the tMSP J1023$+$0038}\label{section:application}

In this Section, we examine whether the pulsar-disk interaction model developed above can account for the observed properties of PSR J1023$+$0038 \citep{Papitto_2019}. In the disk state, PSR J1023$+$0038 emits significantly more high-energy radiation (X-rays, UV, and gamma rays; see \citealt{Patruno_2014, Stappers_2014}) than when it behaves as a rotation-powered pulsar. In addition, it exhibits two similar, bright and broad optical and X-ray pulses per spin period, that extend over half a rotational phase and are separated by $180^\circ$. The model reproduces the increase of the total synchrotron radiation emitted by the pulsar magnetosphere when the disk is present (see Fig.~\ref{fig:skymaps}). According to Figure~\ref{fig:skymaps}, within the family of configurations explored here, the broad double-peaked pulse morphology of PSR J1023$+$0038 is most naturally reproduced by the highest-obliquity case, suggesting that the magnetic moment of PSR J1023$+$0038 might be highly inclined (see also Fig.~\ref{fig:obs_params}, bottom panel). We therefore restrict our analysis to the $\chi=75^\circ$ configuration in the following.

\begin{figure}[bthp]
\centering 
\includegraphics[width=\columnwidth, keepaspectratio]{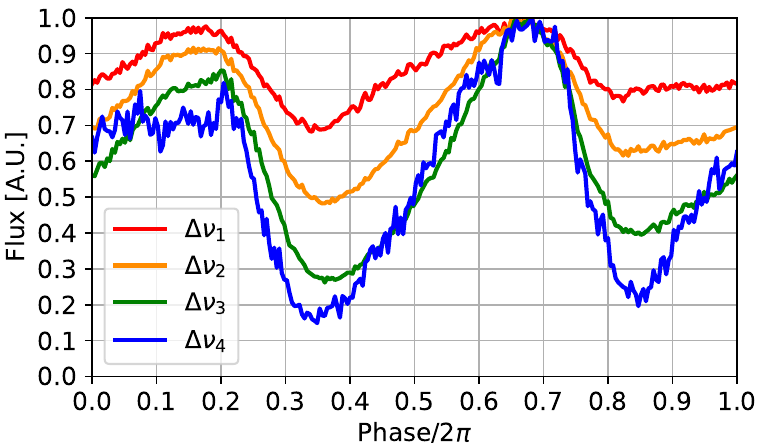}
\caption{Light curves obtained by integrating the emission over four logarithmically spaced frequency bins ($\Delta \nu_1$, $\Delta \nu_2$, $\Delta \nu_3$, and $\Delta \nu_4$), and computed for $\alpha=46.4^\circ$, the estimated viewing angle of PSR J1023$+$0038. Each curve is normalized to its own maximum value for visualization purposes. Data are averaged over $2\,P_{\rm spin}$ after convergence.}\label{fig:energy_bins}
\end{figure}

 We report a slight difference in the peak amplitudes (Fig.~\ref{fig:obs_params}, bottom panel), consistent with observations \citep{Papitto_2019}. Moreover, observed pulsed fractions \citep{Papitto_2019} of $\sim 1\%$ in the optical and $\sim 10\%$ in the X-ray band (i.e., lower than the pulsed fraction typically measured in the X-ray band for isolated pulsars) reflect the presence of a strong continuous component. The numerical results also predict a stronger continuous component than that expected from isolated pulsars (Fig.~\ref{fig:obs_params}, middle panel), although they cannot be directly compared with observations, as our model inherently lacks several sources of continuous emission. In particular, most of the optical flux is produced by the irradiation of the disk or donor star, which are not included in our model. More generally, since the PIC method is well suited for low-density, collisionless environments, the disk is treated as a boundary condition, which prevents us from capturing its emission. Moreover, the thermal radiation, which may contribute significantly, especially from the disk, is neglected. The potential presence of a compact jet in such systems \citep{Baglio_2023} could also provide an additional source of continuous emission. Finally, this study focuses exclusively on synchrotron emission, although such systems may also produce inverse-Compton emission resulting from the interaction between the pulsar wind and photons from the disk or companion. It is worth noting that, given the particle acceleration and photon densities revealed in our simulations, synchrotron self-Compton processes could also be expected.

Although our simulations do not resolve the optical or X-ray bands due to the necessary rescaling of particle-in-cell simulations (see Section~\ref{subsection:initial_setup}), we can nevertheless divide the energy range into different bins to assess the energy dependence of our results. Figure~\ref{fig:energy_bins} presents the emitted synchrotron flux, for $\chi=75^\circ$ and $\alpha=46.4^\circ$ (corresponding to the viewing angle of the system, estimated from optical photometry; \citealt{Stringer_2021}), as a function of the rotation phase, for four logarithmically equally spaced energy bins. Each curve is normalized by its own maximum value. We predict an increase in the pulsed fraction with energy. As mentioned at the beginning of Section~\ref{section:application}, observations show that the pulsed fraction is higher in X-rays than in the optical band. However, within the X-ray band, the source exhibits a constant pulsed fraction with energy ($0.5-50$ keV, see Fig. 13 of \citealt{Papitto_2019}). The total observed emission is made of the sum of the neutron star emission and of the unpulsed emission from other regions (e.g. the disk, the companion star), especially in the optical band. Since our model does not include the dominant optical continuum-emitting components, and the energy ranges are rescaled following the PIC approach, a direct comparison between our predictions and these observations is not possible at this stage.

The pulses remain phase-aligned across the different energy bands, and therefore we do not reproduce the $150\mu$s lag detected between the optical and X-ray pulses. This discrepancy with observations likely arises from the limited extent of our simulation domain, which only covers the inner regions of the wind (up to $5.6\,R_{\rm LC}$) and therefore does not capture particle cooling occurring at larger distances.

\begin{figure}[bthp]
\centering 
\includegraphics[width=\columnwidth, keepaspectratio]{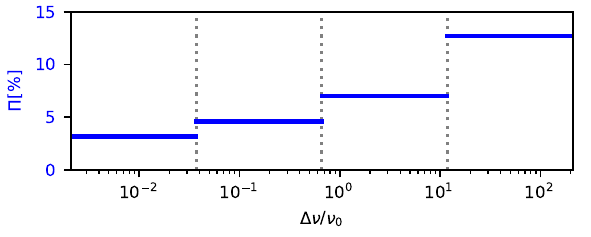}
\caption{Dependence of the polarization degree on frequency for the pulsar-disk case, with $\chi=75^\circ$ and $\alpha=46.4^\circ$ (i.e., the estimated viewing angle for J1023$+$0038). The phase-averaged polarization degree is computed for four logarithmically spaced frequency bins $\Delta \nu/\nu_0$, where $\nu_0=3eB_\star/4\pi m_ec$. Data are averaged over $2\,P_{\rm spin}$ after convergence.}\label{fig:DP_energy}
\end{figure}

The synchrotron radiation produced in our model is linearly polarized, although the presence of the disk reduces the polarization compared to the isolated case, particularly in regions of plasma accumulation. In addition, Figure~\ref{fig:DP_energy} shows that the polarization degree increases noticeably with energy. Observations indicate that the pulsed and polarized fluxes are closely aligned across the spectrum, with polarization degrees of $(1.4\pm0.04)\%$ in the optical band and $(12\pm3)\%$ in the X-rays \citep{Baglio_2025}, consistent with our predictions. Finally, the modeled polarization angle oscillates across the pulsar phase (Fig.~\ref{fig:AP}), similarly to what is reported from observations \citep{Baglio_2025}. 

The main observational features successfully reproduced by the model can be summarized as follows: 
\begin{itemize}
    \item a brighter synchrotron emission from the pulsar magnetosphere in the presence of the disk, compared to when the disk is absent;
    \item two similar, bright pulses,  each lasting about half a spin period and separated by about half a spin period;
    \item a reduced pulsed fraction compared to isolated systems, due to a strong continuous component;
    \item the depolarization of the synchrotron emission compared to isolated systems, the increase of the polarization degree with energy, and a phase-dependent polarization angle.
\end{itemize}
The main remaining discrepancies are:
\begin{itemize}
    \item no phase lag appears across energy bands within the current simulation domain, contrary to the phase lag observed between the optical and X-ray bands;
    \item the continuous component is not fully modeled, owing to the absence of disk emission and additional emission channels.
\end{itemize}

\section{Conclusions}\label{section:conclusion}

We present the first particle-in-cell simulations of a pulsar wind interacting with a surrounding disk, aimed at reproducing the optical and X-ray pulsed emission from the canonical transitional millisecond pulsar PSR J1023$+$0038. We model the resulting rearrangement of the magnetosphere, its associated kinetic effects, and the synchrotron light curves and polarization properties expected from such systems. 

We find a substantial increase in the total luminosity relative to isolated pulsars, driven by increased plasma density between the star and the disk, magnetic field amplification, and enhanced plasma isotropization and particle acceleration. Importantly, we show that the light curves can exhibit either one or two broad peaks per spin period, depending on the pulsar's magnetic obliquity, for a fixed disk aspect ratio ($h/R$). Low magnetic obliquities ($\chi \leq \tan^{-1}(h/R)$) are associated with a single peak, whereas high magnetic obliquities ($\chi \geq \tan^{-1}(h/R)$) result in two peaks separated by half a spin period. At intermediate obliquities, the light-curve morphology evolves progressively from one regime to the other. Exploring a range of disk aspect ratios and distances would provide valuable confirmation of these results. Our highest-obliquity simulation most closely reproduces the observed pulsed emission of PSR J1023$+$0038, featuring two broad peaks separated by $180^\circ$. Expanding the sample of tMSPs will be crucial to determine whether all sources should consistently emit two-peaked pulsed radiation, in which case - according to our model - a high magnetic obliquity is required to reach the subluminous disk state, or whether some systems may instead exhibit single-peaked light curves, indicative of low magnetic obliquities. 

Our model successfully reproduces most of the observed features of PSR J1023$+$0038 in the X-ray high mode, as reported in \citet{Papitto_2019} and \citet{Baglio_2025}. 
Two discrepancies nevertheless remain, concerning the phase shift of the pulsations across energy bands and the strength of the continuous component; these are likely attributable to intrinsic limitations of the current modeling approach. Furthermore, the observed energy dependence of the pulsed fraction, in particular its approximately constant value across the X-ray band, cannot yet be assessed within the present framework.

To conclude, our results confirm a synchrotron origin of the optical and X-ray pulsations, supporting a scenario in which they are generated by the interaction of the pulsar's striped wind and the inner edge of the accretion disk, situated a few light-cylinder radii from the pulsar. Upcoming high-statistics polarimetric observations, together with the discovery of new tMSP sources, will provide valuable constraints on emission models.

\begin{acknowledgements}
We thank R. Mignon-Risse and A. Veledina for insightful discussions, and D.~F. Torres for useful comments on the manuscript. This project has received funding from the European Research Council (ERC) under the European Union's Horizon 2020 research and innovation program (Grant Agreement No. 863412). Computing resources were provided by TGCC under the allocation A0150407669 made by GENCI. A.P. and R.L.P. were supported by INAF Research Grant AF2022 (FANS, PI: Papitto) and AF2024 (PULSE-X, PI: Papitto), the Italian Ministry of University and Research (PRIN MUR 2020, Grant 2020BRP57Z, GEMS, PI: Astone), and Fondazione Cariplo/Cassa Depositi e Prestiti (Grant 2023-2560, PI: Papitto).
\end{acknowledgements}

\bibliographystyle{aa}
\bibliography{main}

\end{document}